# A Storage Ring proton Electric Dipole Moment experiment: most sensitive experiment to CP-violation beyond the Standard Model

Yannis K. Semertzidis
*BNL, Physics Dept., Upton, NY 11973-5000, USA*

for the Storage Ring EDM Collaboration

The Storage Ring EDM Collaboration is completing a comprehensive proposal to DOE for a sensitive proton EDM experiment at the $10^{-29}$ $e$·cm level. This involves building an electric ring of radius R~40 m, with bending provided by stainless steel plates separated by 3 cm gap, providing an electric field of 10 MV/m. This ring can store protons of 0.7 GeV/c momentum, also known as "magic" due to the unique property that the spin and momentum vectors precess at the same rate in any electric field. If the spin is kept longitudinal for most of the duration of the storage time, the radial E-field acts on the proton electric dipole moment and can cause a measureable vertical spin precession. Studies on spin coherence time, on the polarimeter system, and the electric field show that we can reach the experimental goals. A plan on the relative beam position monitors uses currently available technology and applies it in an accelerator environment. Some R&D is needed to establish the feasibility of the plan. Prospects of further improvement with a future upgrade for another order of magnitude in the EDM sensitivity are also laid out.

## 1. Introduction

The presence of a non-zero electric dipole moment (EDM) in a fundamental particle with spin will point to violation of both P-parity and T-time symmetries. Through CPT conservation (with C-charge symmetry), when T is violated, CP is also violated, which, according to A. Sakharov [1], is one of the requirements for an initially symmetric universe to develop to the matter-dominated universe we observe today. Even though CP-violation has been detected in the kaon system at an experiment at BNL [2], it is explained with one CP-violating phase in the CKM matrix [3] in the electro-weak (EW) sector of the standard model (SM). This phase is not nearly enough to explain the ratio of the observed number of baryons over the number of photons in our universe. We observe this ratio to be

$$\eta = \frac{n_B}{n_\gamma} \approx 10^{-10} \tag{1}$$

whereas from the SM we expect this number to be ~$10^{-18}$, some 8 orders of magnitude smaller. Even though the CP-violation accommodated in the EW part of the SM is not adequate to explain the observed asymmetry in the universe, the Lagrangian of the strong interactions makes an a priori large prediction for the EDM in hadronic systems; on the order of $\vartheta \times 10^{-16}$ $e$·cm for the neutron EDM, with $\vartheta$ a QCD parameter of order 1. The limit on the strong CP-violating, so-called $\vartheta$-QCD parameter comes from the limits on the neutron EDM [4] of $|d_n|<2.9\times10^{-26}$ $e$·cm (90% C.L.) resulting to $\vartheta<10^{-10}$. Other potential sources of CP-violation that can be manifested in the neutron or other hadronic EDMs are: SUSY, Multi-Higgs, Left-Right Symmetric, etc. Therefore, deciphering the source of a non-zero EDM measurement of any hadronic system will actually require measuring the EDM of more than one, preferably simple, hadronic system: proton, neutron, deuteron, etc. [5].

A particle with spin has two possibilities for the direction of the EDM (polar) vector: the EDM will be either parallel or anti-parallel to the magnetic dipole moment (axial vector). If the probability for either direction is the same the total EDM expectation value is zero. However, if one direction is preferred, then it clearly violates both the P and T-symmetries since one vector is polar (EDM) and the other axial (magnetic dipole moment).

The timeline of the neutron (dark), and electron (red) EDM experimental limits as well as the various theoretical models that could give rise to an EDM is shown in Fig. 1. In the figure we also indicate the neutron, proton, deuteron and electron EDM experimental targets. If those goals are met they will test the Electroweak Baryogenesis hypothesis. Clearly reaching down to and probing the Electroweak Baryogenesis models is of fundamental importance in our effort of understanding the origin of the universe. The SM predictions of the electron EDM is in the range of $d_e$~$10^{-41}$ $e$·cm and that of the neutron at the $d_n$~$10^{-31}$ $e$·cm level, well below the present experimental capabilities.

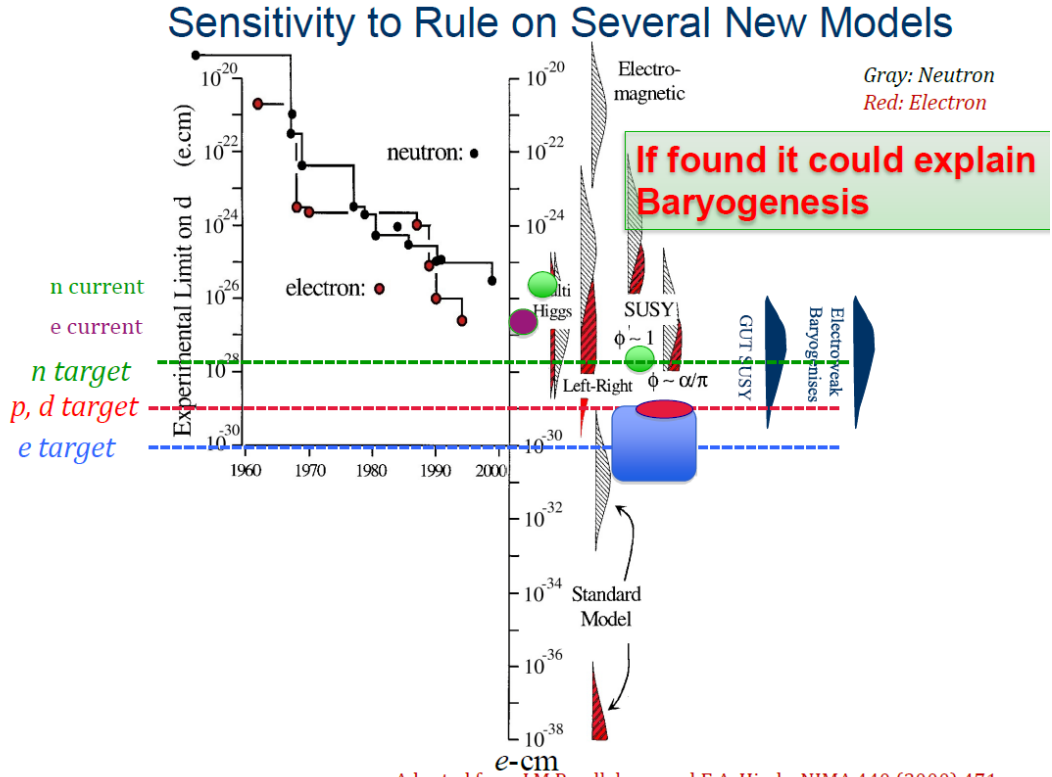

Figure 1: The experimental limits of the neutron and electron as a function of year as well as theoretical models motivating those searches are shown here. The neutron, proton, deuteron, and electron target experiments plan to have results within the next ten years at which time they will severely test the Electroweak Baryogenesis hypothesis.

The measurement of the neutron EDM requires the application of a magnetic field and of an electric field and observing the change in the interaction energy with the flip of the E-field. The spin precession rate for a particle at rest is given by:

$$\frac{d\vec{s}}{dt} = \vec{\mu} \times \vec{B} + \vec{d} \times \vec{E} \qquad (2)$$

The magnetic field is required to keep the neutrons polarized and precessing at a specific frequency. The electric field is flipped with the change in the precession rate measured to yield the EDM value:

$$d = \frac{\hbar(\omega_1 - \omega_2)}{4E} \qquad (3)$$

meaning we would get a spin precession of ~6 nrad/s for $d_n=10^{-29} e\cdot$cm in an electric field of 100 kV/cm, i.e. one turn per 30 years, which is quite small. To get an idea of the smallness of the EDM level for $d_n=10^{-29} e\cdot$cm, if we blow-up the neutron to become as large as the sun, the center of gravity of the two charges (positive and negative) will be split along the direction of the sun's rotation axis by one tenth of a micron!

The statistical sensitivity of the EDM depends on the particulars of the experimental measurement cycle time. When the measurement cycle time is longer than the time required preparing a new pulse of polarized particles, the statistical error is given by

$$\sigma_d = \frac{b\hbar}{EPA\sqrt{Nf\tau\,T}} \qquad (4)$$

with *P*, *A* the polarization and analyzing power respectively, *N* the number of injected particles per storage time, *f* the detection efficiency of the detector, *τ* the spin coherence time and *T* the lifetime of the experiment. *b* is a factor of order 1 depending on the particulars of the storage time optimization.

A proton EDM at the $10^{-29}e$·cm level will improve the $\vartheta$-QCD sensitivity by more than three orders of magnitude to below $0.3\times10^{-13}$, and it will have sensitivity to new contact interactions at the 3000 TeV level. For SUSY-like new physics its sensitivity level is

$$d_p \approx 10^{-24} e\cdot\text{cm} \, \sin\delta \times \left(\frac{1\,\text{TeV}}{M_{SUSY}}\right)^2 \quad (5)$$

with *M* the SUSY-like mass scale of the new physics and *δ* the value of the CP-violating phase in the model. The mass range is at the 300 GeV or, if there is new physics at the LHC scale, the CP-violating phase sensitivity is at the 0.1 μrad level. The relevant mass scale for the LHC-scale SUSY is the neutralino mass.

## 2. Experimental Method

The storage ring EDM method was studied in detail for the muon [6], the deuteron [7] and the proton [8] cases. We've estimated we could reach $10^{-24}e$·cm for the muon running for several years at J-PARC; $10^{-29}e$·cm for the deuteron running several years at BNL, and $10^{-29}e$·cm for the proton within a couple of years at BNL. The BNL PAC was enthusiastic on the physics reach of the experiment and recommended technical reviews to establish the feasibility of the experiment. We had two, very successful, technical reviews one in December 2009 and one in March 2011. Currently we are finishing up the proposal to DOE to build the largest electric ring in the world (*R*~40 m), storing $4\times10^{10}$ polarized protons at a time for ~$10^3$s in both the clock-wise (CW) and counter-clock-wise (CCW) directions.

The proton EDM ring differs from both the deuteron and muon rings we proposed in the past in that we require only electric fields, while we proposed using a combination of magnetic and electric fields for the deuteron and muon rings. We believe using only electric fields simplifies the experiment and thus we suggest starting with the proton EDM experiment.

From eq. (2) above it is clear that one needs to apply an electric field to probe the particle EDM. Applying an electric field on a charged particle is cumbersome since it will drive it away from the storage region and most likely into the walls of the plates generating the electric field. However, if the particle has the correct energy and the ring has the correct focusing characteristics the particle will be stored indefinitely. For a particle in motion the spin precession rate with respect to the momentum precession rate is given by

$$\vec{\omega}_a = -\frac{q}{m}\left\{\left[a-\left(\frac{m}{p}\right)^2\right]\frac{\vec{\beta}\times\vec{E}}{c}\right\} \quad (6)$$

with *q*, m the charge and mass of the proton respectively, *a*=1.792847357(23) the proton anomalous magnetic moment, *p* its momentum and *β* the proton speed in units of *c*. It may look counter-intuitive to have a spin precession in the presence of only electric fields (E). This is only so because a moving particle will feel both a magnetic and electric field in its own rest frame due to Lorentz transformation of the lab E-field. For a positive anomalous magnetic moment (*a*) it is also possible to make the right hand side of eq. (6) equal to zero for

$$p = \frac{m}{\sqrt{a}} \approx \frac{0.938}{\sqrt{1.79}} \approx 0.7\,\text{GeV/c} \quad (7)$$

with the numerical example being valid for the proton. Clearly this value is independent of the E-field value. The so-called "magic" momentum is different for different particles. It is 3.1 GeV/c for the muon, the value used for the last muon g-2 experiment at CERN [9] and the muon g-2 experiment at BNL [10]. In an all-electric ring running at the "magic" momentum means that the spin and momentum vectors precess horizontally at the same rate, as shown in Fig. 2 below.

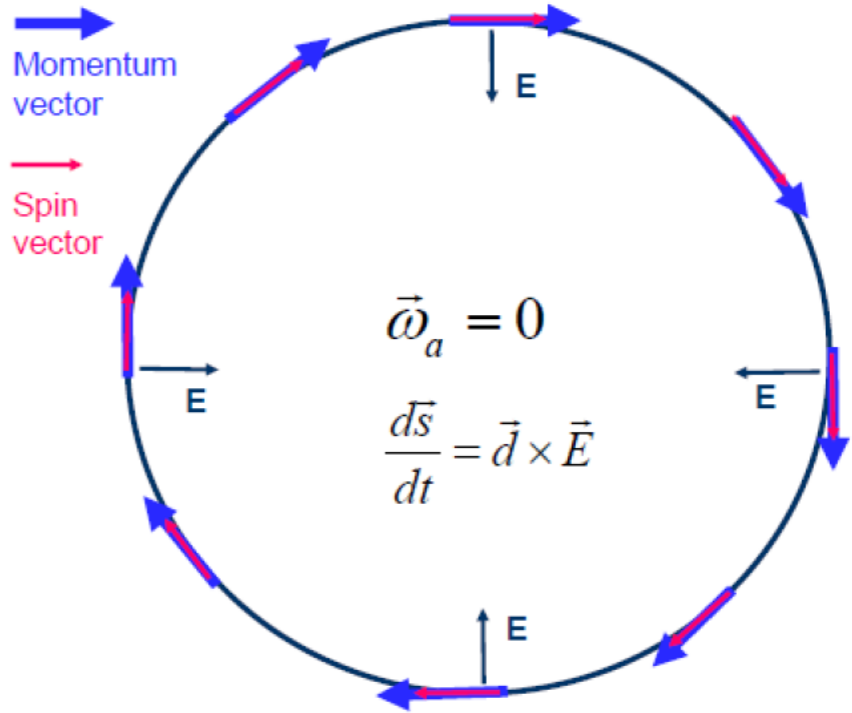

Figure 2: Top view of the proton momentum and spin vectors in an all-electric storage ring. The spin and momentum vectors precess at the same rate horizontally. The radial E-field is always on the right of the particle as it moves around the ring acting on the proton EDM for the duration of the storage time. The proton spin will optimally rotate out of the plane if the spin orientation is kept longitudinal for the duration of the storage time.

Following injection the average vertical spin direction of the stored beam will be determined by slowly reducing the vertical tune of the storage ring from 0.2 to 0.1 driving the beam into the aperture defining solid carbon target, which is part of the polarimeter detector. Nuclear elastic proton-carbon scattering is used to determine the proton transverse spin components as a function of time. The average analyzing power for scattering angles between 5° and 20° in the lab frame is estimated to be 60% with a collection efficiency of more than 0.5% [8].

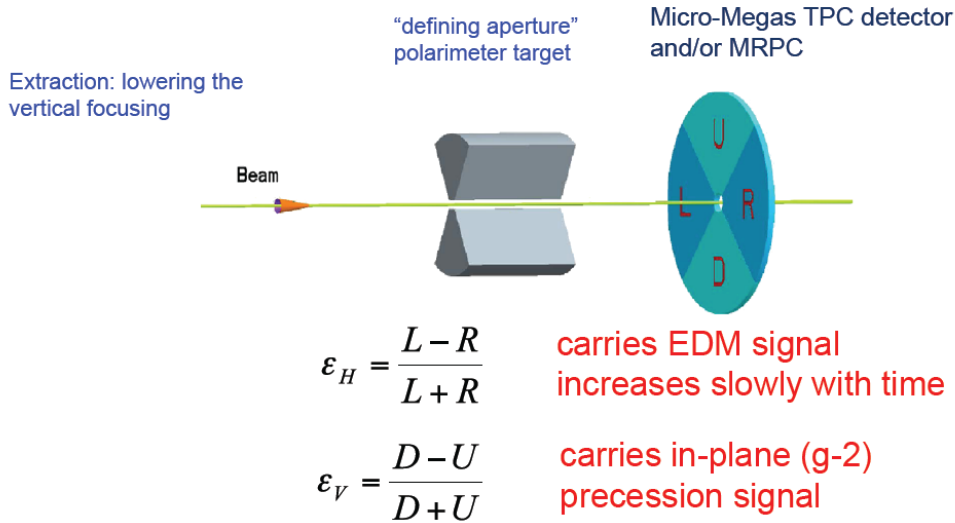

Figure 3: The polarimeter system includes an aperture defining solid carbon target and a detector about 1m away from it. Nuclear elastic scattering of protons with carbon have a large analyzing power of the proton transverse spin components.

The change of the left counter rate minus the right counter rate over the sum as a function of time indicates the presence of an EDM signal, as shown in Fig. 4. The statistical sensitivity is estimated using eq. (4) with the value of $b=2$ when a uniform data rate is used and $b=1.4$ when the data rate is enhanced by a factor of four at the start and end of the storage time. The values of the parameters used are: $P=0.8$, $A=0.6$ the polarization and analyzing power respectively, $N=4\times10^{10}$ the number of injected particles per storage time, $f=0.5\%$ the detection efficiency of the detector, $\tau=10^3$s the spin coherence time and $T=4\times10^7$ s (equivalent to four calendar years) the lifetime of the experiment. The estimated statistical error is $1.1\times10^{-29}$ $e\cdot$cm per year (defined as $10^7$ s).

The spin tracking indicates that the spin coherence time is long enough to accommodate beam storage for $10^3$ s. Estimations of the spin coherence time using stochastic cooling indicate that it can be prolonged by a couple of orders of magnitude making possible another order of magnitude improvement in the statistical sensitivity. The stochastic cooling of the beam would also allow further lowering the vertical tune and thus enhancing the beam position monitors sensitivity to the radial B-field and thus an order of magnitude improvement over the $10^{-29}$ $e\cdot$cm sensitivity seems to be within reach. More studies are needed to establish this improvement in sensitivity.

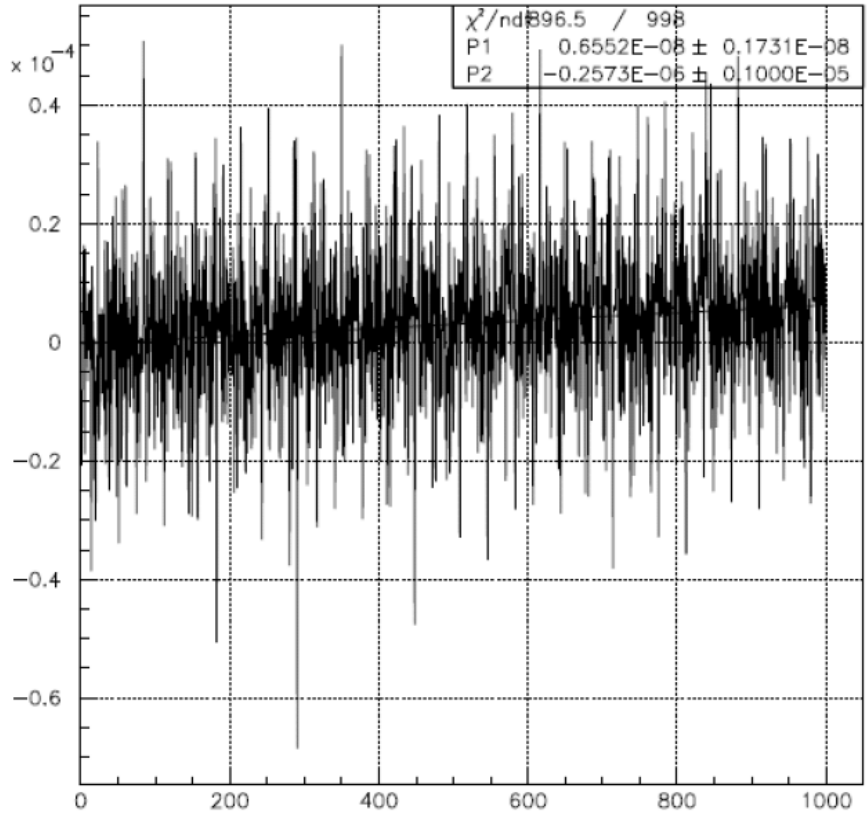

Figure 4: Monte Carlo (M.C.) data showing the (L-R)/(L+R) ratio as a function of time [s] in case of a vertical spin developing as a function of time. The M.C. results are in agreement with eq. (8).

The power of this method is clearly evident when one considers the beam intensities of $4\times10^{10}$, highly polarized (>80%) protons, that can be stored for about $10^3$ s in the presence of strong electric fields of order 10 MV/m. The detection efficiency is >0.5% ensuring the statistical sensitivity goal within a year or two of running. Spin tracking simulations show that the particles within the admittance of the ring will have the required spin coherence time even without using sextupoles. That means that the second order effects in an all-electric ring are small and adequate for the needs of the experiment.

The beam dynamics of an all-electric ring can be inferred from the study given in ref. [11]. The radial excursion due to momentum dispersion, for a homogenous ring and cylindrical geometry for the E-field plates:

$$x_D = D_x \lambda_p = R_0 \left( \frac{dp}{p_0} - \frac{x}{R_0} \right), \quad D_x = R_0$$

$$v_{x0}^2 = 1 + \frac{1}{\gamma_0^2}, \quad \frac{df}{f_0} = -\left( \frac{dp}{p_0} - \frac{x}{R_0} \right)$$

(8)

The acceptance of the ring has been estimated using tracking and is about 6 mm mrad, good enough to admit protons with dp/p~2×10⁻⁴ and a horizontal angle of ~0.5 mrad. Accelerator studies at BNL showed that the beam intensity and phase-space requirements could be met by scraping the available polarized proton beam in both the horizontal and vertical directions.

The major systematic error is a net radial B-field around the ring, which needs to be shielded from, using static and dynamic methods, with a combined shielding factor of $10^7$-$10^9$. Owing to the details of the beam and spin dynamics in storage rings it turns out that the geometrical phase (GP) effects will be below the statistical sensitivity of the experiment. The construction specs of the ring are limited by the GP effects and are well within the customary requirements of building a storage ring and therefore those are not expected to be a major issue. Polarimeter systematic errors, related to beam motion on the target are also going to be well below our sensitivity level as is shown from our runs at the COSY ring in Jülich/Germany. A long paper on the polarimeter systematic error tests has recently been accepted for publication in NIMA.

## 3. Electric field development

The electric field goal is to apply ±150 kV on metal plates separated by 3 cm. The electric field plates look very much like the traditional electro-static separators used previously at BNL and FNAL, see Table I, which were designed more than 10 to 20 years earlier. The applied voltage on the plates is similar to those applied in past electro-static separators, however, the plate distance is smaller in the proton EDM experiment. We expect to be able to raise the applied electric field by specially treating the metal surfaces with high-pressure water rinsing [12], see Fig. 5.

Table I: Large scale E-field electrodes

| Parameter | Tevatron pbar-p Separators | BNL K-pi Separators | Proton EDM E-field plates |
|---|---|---|---|
| Length | 2.6 m | 4.5 m | 3 m |
| Gap | 5 cm | 10 cm | 3 cm |
| Height | 0.2 m | 0.4 m | 0.2 m |
| Number | 24 | 2 | 84 |
| Maximum High Voltage | ±180 kV | ±200 kV | ±150 kV |

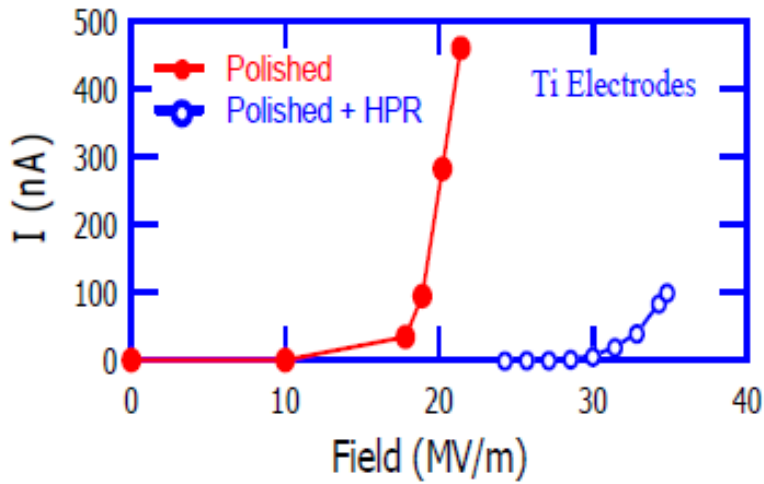

Figure 5: The field emission as a function of the applied E-field between two parallel plates before (red) and after (blue) high pressure water rinsing, from [12].

## 4. Radial B-field systematic error and B-field shielding

The only major systematic error is the average radial B-field around the ring. The average radial B-field needs to be below 1nG with 1 Hz BW. The average radial B-field will vertically split the counter-rotating beams by an amount that depends on the vertical focusing of the ring. We plan using a weak vertical tune of order 0.1 (the muon g-2 experiment at BNL run as low as at 0.12 for significant part of data taking) and modulate that tune by 10% at a frequency of our choice, i.e. the functional form of the tune will be

$$m = m_0 \left(1 + 0.1 \times \cos(\omega t + \varphi)\right) \approx 0.1 + 0.01 \times \cos(\omega t + \varphi) \quad (9)$$

The splitting between the counter-rotating beams will produce an oscillating radial magnetic field in the storage plane which we plan [8] to observe using SQUIDs. We are following a similar method as the one tested in the lab [13] where they have achieved the required sensitivity in the lab. The frequency of modulation is going to be in the 0.1-10 kHz range. The high frequency magnetic field noise originating from the beam is going to be attenuated by a thin metal shield between the SQUIDS and the beam. The outside B-field noise is going to be shielded by a number of concentric mu-metal shields, as well as 1 cm thick aluminum cylinder. Fig. 6 shows the shielding achieved in several systems as well as the requirement for our experiment (red oval). It is evident that we can achieve the required shielding factor.

The ring itself will be shielded by three to four mu-metal shields reducing the outside B-field by a factor up to $10^5$. The rest of the needed shielding factor would need to come from active feedback: 1) using Helmholtz coils in the ring tunnel, and 2) inside the beam-tube again using Helmholtz coils. The average horizontal magnetic field component due to the earth's magnetic field is of order 0.2 G. The average radial magnetic field in the ring due to the earth's magnetic field should be very close to zero, but here we'll assume it 10 mG. At every second the average radial B-field needs to be below ~1 nG, i.e. there is a shielding factor requirement of $10^7$. We expect to get $10^4$-$10^5$ from a passive shield [15] and the rest from the active shield.

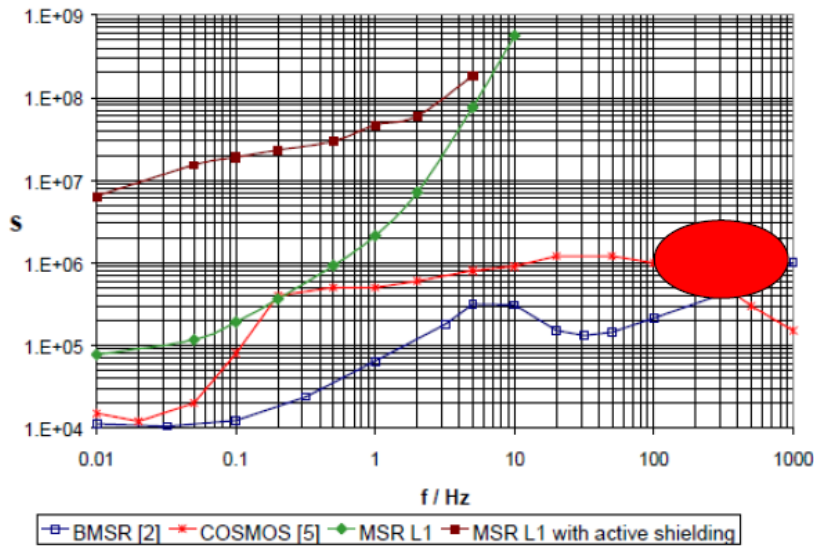

Figure 6: The shielding factor vs. frequency. The MSR L1 without (green) and L1 with active shielding are the two top-leftmost curves. One of the magnetic shielding layers is a 1 cm thick aluminum tube. The red ellipse shows what (passive) shielding factor we need to achieve at the BPM locations. The BMSR (red) and COSMOS (blue) lines are older achievements, with the references given in [14].

## 5. Similarities and differences between the neutron and "magic" momentum proton EDM experiments

There are similarities and differences between the neutron and proton EDM experiments. They both require large number of polarized particles stored for significant times and probed by a large electric field. They both have a similar systematic error, originating from spurious magnetic fields. They are given in table II below:

Table II: Similarities and differences between the neutron and proton EDM experiments.

| nEDM | pEDM |
|---|---|
| Apply a weak magnetic field (~1μG) and a strong E-field (~50 kV/cm) between plates ~10cm apart. | Eliminate all B-fields. Apply a strong E-field (~100 kV/cm) between plates 3 cm apart. |
| EDM signal: Look for a precession frequency change when the E-field changes direction. | EDM signal: Look for a change in the vertical spin component as a function of storage time. |
| Major systematic errors: B-field stability vs. time; E and B-field alignment, dB/dz < 0.1 μG/cm (Geometrical phase). | Major systematic errors: a non-zero radial B-field. It needs to be reduced to below 1 nG with 1 Hz BW using passive and active shielding. Use the counter-rotating beam splitting to sense the radial B-field. The Geometrical Phase effect is below the statistical sensitivity of the experiment for commonly achievable plate tolerances. |
| Need ultra-cold neutrons. Current density 5n/cc, aiming for 150n/cc and ~5 lt storage volume. | Cold protons ($dp/p \leq 2 \times 10^{-4}$). Beam intensity $10^{10}$ highly polarized in both CW and CCW directions. |

## 6. Conclusions

The storage ring EDM collaboration is preparing a proposal for a proton EDM experiment with $1.1 \times 10^{-29} e \cdot$cm per year sensitivity. The method is using counter-rotating protons at their "magic" momentum stored in an all-electric storage ring of radius $R \sim 40$ m, and an E-field of 10 MV/m between plates at 3 cm distance. The (only) major background of the experiment is a net radial B-field around the ring. The vertical separation between the counter-rotating beams depends on the vertical focusing. The vertical tune is kept at 0.1 and is modulated at ~10% of itself to make it easier to detect the vertical separation. Our beam position monitors are designed to have the needed sensitivity to place a strict limit on this background using currently available technology. We plan to test it in the RHIC accelerator for proof of principle that the method can be applied in an accelerator environment. All other major issues of the experiment are at a very advanced stage and therefore we believe we are ready for a CD0 stage.